\documentclass[prl,preprint,superscriptaddress]{revtex4}
\usepackage{dcolumn}
\usepackage{mathtools}
\usepackage{amssymb}
\usepackage{bm}
\usepackage{pifont}
\usepackage{psfrag}
\usepackage{epstopdf}
\usepackage{wasysym}
\usepackage{amsmath}
\usepackage{tipa}
\usepackage{hyperref}
\usepackage{setspace}
\hypersetup{
     colorlinks = true,
     linkcolor = blue,
     anchorcolor = blue,
     citecolor = blue,
     filecolor = blue,
     urlcolor = blue
     }
\usepackage[usenames]{color}
\renewcommand{\footnoterule}{%
  \hrule width \textwidth height 1pt
  \kern 2pt
}
\begin{document}

\title{Tuning the performance of a micrometer-sized Stirling engine through reservoir engineering}
\author{Niloyendu Roy \footnote[1]{ $\space$ Corresponding author. email: \href{mailto: niloycrj@gmail.com}{niloycrj@gmail.com} }}
\affiliation{Chemistry and Physics of Materials Unit, Jawaharlal Nehru Centre for Advanced Scientific Research, Jakkur, Bangalore - 560064, INDIA}
\author{Nathan Leroux}
\affiliation{Unit\'e Mixte de Physique CNRS/Thales, 91767 Palaiseau, France}
\author{A K Sood}
\affiliation{Department of Physics, Indian Institute of Science, Bangalore- 560012, INDIA}
\affiliation{International Centre for Materials Science, Jawaharlal Nehru Centre for Advanced Scientific Research, Jakkur, Bangalore - 560064, INDIA}
\author{Rajesh Ganapathy \footnote[2]{ $\space$ Corresponding author. email: \href{mailto: rajeshg@jncasr.ac.in}{rajeshg@jncasr.ac.in}}}
\affiliation{International Centre for Materials Science, Jawaharlal Nehru Centre for Advanced Scientific Research, Jakkur, Bangalore - 560064, INDIA}
\affiliation{School of Advanced Materials (SAMat), Jawaharlal Nehru Centre for Advanced Scientific Research, Jakkur, Bangalore - 560064, INDIA}

\date{\today}

\begin{abstract}
\textbf{Colloidal heat engines are paradigmatic models to understand the conversion of heat into work in a noisy environment - a domain where biological and synthetic nano/micro machines function. While the operation of these engines across thermal baths is well-understood, how they function across baths with noise statistics that is non-Gaussian and also lacks memory, the simplest departure from equilibrium, remains unclear. Here we quantified the performance of a colloidal Stirling engine operating between an engineered \textit{memoryless} non-Gaussian bath and a Gaussian one. In the quasistatic limit, the non-Gaussian engine functioned like an equilibrium one as predicted by theory. On increasing the operating speed, due to the nature of noise statistics, the onset of irreversibility for the non-Gaussian engine preceded its thermal counterpart and thus shifted the operating speed at which power is maximum. The performance of nano/micro machines can be tuned by altering only the nature of reservoir noise statistics.}
\end{abstract}
\maketitle

%For the non-Gaussian Stirling engine $\langle W_{cyc} \rangle = -0.0059 \pm 0.001\text{ }k_BT_{eff}^C$ and $W_{\infty} = -0.0065 \text{ }k_BT_{eff}^C$ and for the thermal one $\langle W_{cyc} \rangle = -0.0042 \pm 0.0013\text{ }k_BT_{eff}^C$ and $W_{\infty} = -0.0054 \text{ }k_BT_{eff}^C$.

%To obtain statistically meaningful quantities, the number of Stirling cycles, $N$, executed was in excess of 250 cycles for the large $\tau$ and was increased to almost 2000 cycles for the smallest $\tau$ studied.

Experimental advances in nano/micro manipulation has made feasible the realization of mesoscale heat engines with only a single atom \cite{rossnagel2016single} or colloidal particle \cite{blickle2012realization,quinto2014microscopic,martinez2016brownian,ciliberto2017experiments,martinez2017colloidal} as the working fluid. Even while the functioning of these engines is strongly influenced by fluctuations in the local environment with parameters like work and efficiency becoming stochastic quantities, when operating between equilibrium heat baths, their cycle-averaged performance mirrors their macroscopic counterparts and standard thermodynamic relations apply \cite{sekimoto1998langevin,sekimoto2010stochastic,esposito2010efficiency,seifert2012stochastic,verley2014unlikely,rana2014single}. Recently, Krishnamurthy et al. \cite{krishnamurthy2016micrometre} experimentally realized an \textit{active} stochastic heat engine by replacing the isothermal branches of a Stirling cycle with isoactive ones. Here, a colloidal particle in a time-varying optical potential was periodically cycled across two bacterial reservoirs characterized by different levels of activity. Unlike in equilibrium thermal baths where the displacement distribution  of the colloid, $\rho(x)$, is a Gaussian, in active reservoirs, it was non-Gaussian and heavy-tailed \cite{krishnamurthy2016micrometre,wu2000particle}. These rare large displacement events resulted in large work output and the efficiency of this active engine was found to surpass equilibrium engines; even those operating between thermal baths with an infinite temperature difference. Since the metabolic activity of the bacteria could not be altered rapidly, this engine was operated only in the quasistatic limit, i.e. for a cycle duration $\tau$ larger than the relaxation time of the colloid. Subsequent theoretical calculations for the $\tau\to\infty$ limit posited that a departure from equilibrium efficiencies requires noise not just with non-Gaussian statistics but also with memory, a feature typical of active baths due to the persistent motion of the particles \cite{zakine2017stochastic}. In fact, when the bath noise is non-Gaussian and white, an effective temperature $T_{eff}$ defined through the variance of $\rho(x)$ is thought to act like a bona fide temperature \cite{zakine2017stochastic,fodor2018non} and engines operating between such baths are expected to perform like equilibrium ones in the quasistatic limit. Whether this similarity persists when $\tau$ is reduced and irreversibility begins to set in is not known and is worth exploring since real heat engines never operate in the quasistatic limit as here their power $P\to 0$. On the experimental front, memoryless non-Gaussian heat baths are yet to be realised and predictions even in the quasistatic limit remain untested.

Here we engineered a \textit{memoryless} non-Gaussian heat bath and then constructed and quantified the functioning of a colloidal Stirling heat engine operating between such a bath and a thermal one for different $\tau$. In the quasistatic limit, the performance of this non-Gaussian engine mirrored a classical Stirling engine operating between thermal/Gaussian baths in agreement with theoretical predictions. Strikingly, due primarily to differences in the noise statistics of the baths, the small $\tau$ behaviour of these engines was quite different. On lowering $\tau$, not only did the distribution of work done per cycle, $\rho(W_{\text{cyc}})$, by the non-Gaussian engine become increasingly negatively skewed, unlike the standard Stirling case where it remained Gaussian, the onset of irreversibility for these two engines was also different. Importantly, we demonstrate that even sans memory, changing the nature of noise statistics of the reservoirs between which an engine operates allows tuning its performance characteristics, specifically, the $\tau$ at which the power goes through a maximum.

\section*{\large Results}

\subsection*{Reservoir engineering by flashing optical traps}
\begin{figure}[tbp]
\includegraphics[width=0.82\textwidth]{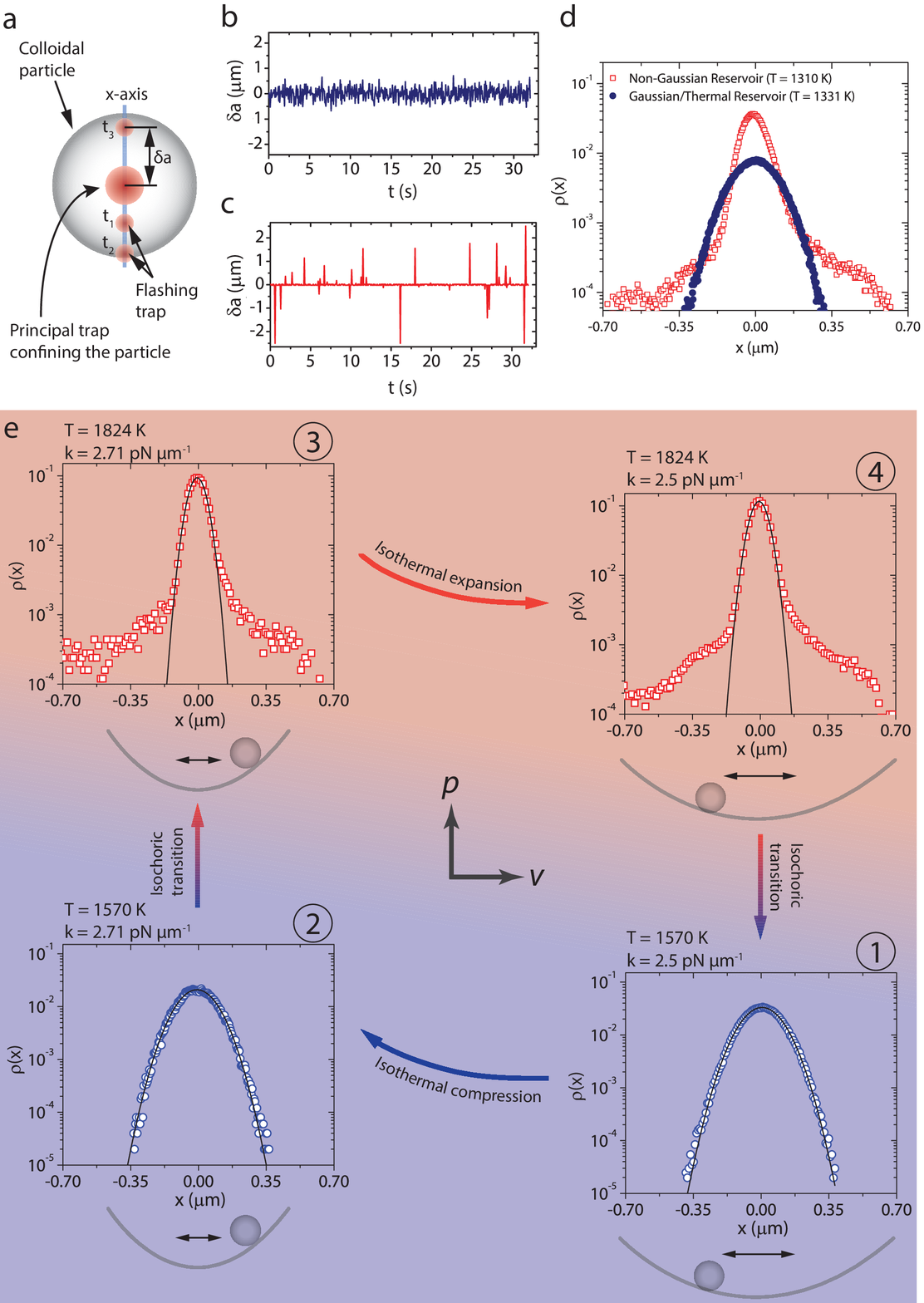}
\caption{\textbf{Experimental realization of a non-Gaussian Stirling heat engine.} \textbf{a} The big red spot represents the primary optical trap and the small red spots represent the secondary flashing optical trap at different time instances $t_1$, $t_2$ and $t_3$. \textbf{b} and \textbf{c} show the distance $\delta a(t)$ from the primary trap at which the secondary trap was flashed as a function of $t$ for engineering a Gaussian and a non-Gaussian reservoir, respectively. \textbf{d} shows the probability distribution of particle displacements, $\rho(x)$, for the engineered Gaussian/thermal (solid blue circles) and the non-Gaussian reservoir (red hollow squares) for a nearly identical $T_{eff}$. \textbf{e} shows a quintessential Stirling cycle between a hot non-Gaussian bath at $T_{eff}^H = 1824$ K and a cold Gaussian reservoir with $T_{eff}^C = 1570$ K. The trap stiffness $k$ is varied linearly in the expansion/compression steps. Having a fixed primary trap and a second flashing optical trap, as opposed to just the latter, prevented the trapped particle from escaping the trap and allowed for long experiments. $\rho(x)$ of the particle measured at the four state points labeled \textcircled{1} to \textcircled{4} is also shown. The black lines are Gaussian fits.}
\label{Figure1}
\end{figure}

Our experimental scheme for reservoir engineering is elaborated in Figure \ref{Figure1}a. A polystyrene colloidal particle of radius $R = 2.5$ $\mu$m suspended in water is held in a harmonic optical potential, $U = {1\over2}k_1\langle x^2\rangle $, created by tightly focusing a laser beam (1064 nm ALS-IR-5-SF, Azur Light Systems France) through a microscope objective (Leica Plan Apochromat 100X, N.A. 1.4, oil) that is also used for imaging the particle (see methods). Here, $k_1$ is the stiffness of this primary trap, $x$ is the displacement of the colloid from the centre of the optical trap and $\langle\rangle$ denotes an average. At equilibrium, the trap stiffness can be determined through  the equipartition relation  ${1\over2}k_1\langle x^2\rangle =  {1\over2}k_BT$ where $k_B$ is the Boltzmann constant and $T$ is the bath temperature, which in our experiments is fixed at 300 K. As a first step, we attempted to engineer a reservoir that mimicked a thermal bath, i.e. with Gaussian noise statistics, but with a desired $T_{eff}$. To this end, we imposed an additional noise on the colloidal particle along one spatial dimension, here the $x-$axis (Figure \ref{Figure1}a), from a second optical trap of fixed intensity but with a time-dependent centre that was flashed at a distance $\delta a(t)$ away from the primary one (Figure \ref{Figure1}b). This was made possible by using a second laser (Excelsior 1064 nm, Spectra Physics USA) coupled to the microscope through a spatial light modulator (Boulder Nonlinear Systems USA) and the flashing frequency was held fixed at $34$ Hz (see Methods). Earlier reservoir engineering studies wherein the colloidal particle experienced only the potential from the flashing trap found that when $\delta a$ was drawn from a Gaussian distribution, the particle indeed behaved like one in a thermal bath but at a $T_{eff}>T$ and furthermore, when $\delta a(t) < R$, the trap stiffness also remained unaltered. \cite{berut2014energy,chupeau2018thermal}. Here we adhered to the same protocol and further ensured that the peak of the $\delta a$ distribution coincided with the centre of the primary trap. Thus, the effective trap stiffness in our experiments $k = k_1 + k_2$, where $k_2$ is the stiffness of the flashing trap. Like in a thermal bath, $\rho(x)$ of the trapped colloidal particle was a Gaussian (solid circles in Figure \ref{Figure1}d) and its power spectral density (PSD) a Lorentzian, allowing us to determine $k_2$ and hence $T_{eff}$ \cite{chupeau2018thermal} (Supplementary Figure 1 and Supplementary Note 1). For the $\delta a(t)$ profile shown in Figure \ref{Figure1}b, the particle experienced a $T_{eff} = 1331$ K.

Engineering a memoryless non-Gaussian reservoir involved only a small tweak to the manner in which the external noise was imposed on the colloidal particle. The instantaneous $\delta a$ was now drawn randomly from a distribution with zero mean and skew, as before, but with a high kurtosis (see Methods and Supplementary Figure 2). Such a distribution has a narrow central region with heavy tails. The flashing optical trap is thus mostly coincident with the primary trap, thereby confining the particle strongly, and is occasionally positioned a large distance away from the centre leading to a large excursion by the particle (Figure \ref{Figure1}c and Supplementary Movie). The overall noise experienced by the particle is $\delta$-correlated as the thermal and imposed noise are individually $\delta$-correlated. Under the influence of such a noise, the corresponding $\rho(x)$ of the colloidal particle was also non-Gaussian (hollow squares in Figure \ref{Figure1}d). The PSD of the particle could be fit to a Lorentzian over the dynamic range accessible and since all other experimental parameters are held fixed, the roll-off frequency of the PSD was also same as the Gaussian case (Supplementary Figure 3 and Supplementary Note 1). For an appropriate choice of the variance and kurtosis of the $\delta a$ distribution, we could engineer the $T_{eff}$ of the non-Gaussian bath, again defined through the variance of $\rho(x)$, to be nearly identical to that in a Gaussian bath (Figure \ref{Figure1}d).      

\subsection*{Performing a Stirling cycle between engineered reservoirs}

Armed with the capability to engineer reservoirs, we built a colloidal Stirling engine operating between a hot non-Gaussian and a cold Gaussian bath held at temperatures  $T_{eff}^H = 1824$ K and $T_{eff}^C = 1570$ K, respectively. We also compared the performance of this non-Gaussian engine with a standard Stirling engine operating between engineered Gaussian baths with similar effective temperature difference (see Supplementary Note 2). The Stirling cycle we executed with the trapped colloid (Figure \ref{Figure1}e), like in previous studies \cite{blickle2012realization,krishnamurthy2016micrometre,schmiedl2007efficiency}, comprised of an isothermal compression (path \textcircled{1} - \textcircled{2}) and expansion step (path \textcircled{3} - \textcircled{4}) linked by two isochoric transitions (paths \textcircled{2} - \textcircled{3} and \textcircled{4} - \textcircled{1}). In the isothermal compression (expansion) steps, $k$ was increased (decreased) linearly from $k_{min} = $2.5 pN$\mu$m$^{-1}$ to $k_{max} = $ 2.7 pN$\mu$m$^{-1}$ by changing $k_1$ alone. The isochroric transitions were near instantaneous and occurred on millisecond time scales. We exploited the ability to rapidly alter $T_{eff}$ and also the nature of noise statistics through the SLM to explore engine performance over a range of $\tau$ which spanned from 2 s to 32 s (see Methods).  

\subsection*{Elucidating the origins of irreversibility in the non-Gaussian Stirling engine}

\begin{figure}[tbp]
\includegraphics[width=0.85\textwidth]{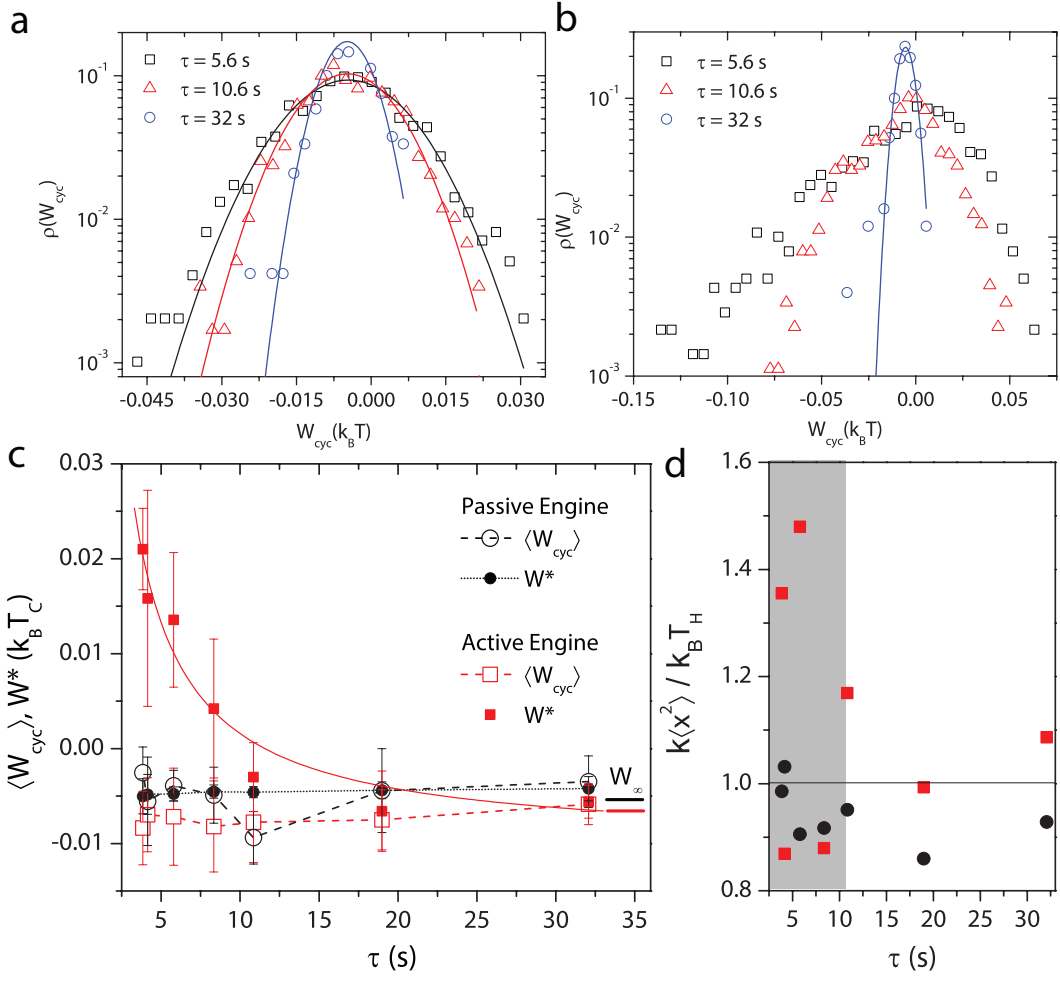}
\caption{\textbf{Buildup of irreversibility in the non-Gaussian Stirling engine at finite $\tau$.} \textbf{a} and \textbf{b} show the probability distribution of work done per cycle $\rho(W_{cyc})$ for the Gaussian and the non-Gaussian engine, respectively, for $\tau = 32$ s (blue circles), $10.6$ s (red triangles) and $5.6$ s (black squares). Solid lines represent corresponding Gaussian fits to the data. \textbf{c} Red hollow and solid squares show the average work done per cycle $\langle W_{cyc} \rangle$ and the most-probable work $W*$, respectively, for the non-Gaussian engine at various $\tau$. The red solid line is a fit to Equation \ref{1}. Black hollow and solid circles show  $\langle W_{cyc} \rangle$ and $W^*$ respectively for the thermal/Gaussian engine. 
At large $\tau$, the experimentally calculated work for these engines agrees with theoretically calculated quasistatic work $W_{\infty}$ indicated by the Red and Black short horizontal lines for the non-Gaussian and Gaussian engine respectively. \textbf{d} Red squares (black circles) represent the ratio $k\langle x^2 \rangle / k_BT_{eff}^H$ calculated at the midpoint of the hot isotherm of the non-Gaussian (Gaussian) engine at various $\tau$. The horizontal line indicates the equilibrium condition, which is strongly violated inside the shaded grey region, in case of the non-Gaussian engine.} 
\label{Figure2}
\end{figure}

The framework of stochastic thermodynamics provides a prescription for calculating thermodynamic quantities like the work, power, and efficiency of mesoscopic machines \cite{sekimoto1998langevin,sekimoto2010stochastic,seifert2012stochastic,schmiedl2007efficiency}. The work done per cycle, $W_{cyc}$, by the particle due to a modulation in the stiffness of the trap is just the change in potential energy and is given by $W_{cyc} = \int_{t_i}^{t_i+\tau} \frac{\partial U}{\partial k}\circ dk \equiv \frac{1}{2}\int_{t_i}^{t_i+\tau} x^2 \circ dk$. Here, the $\circ$ signifies that the product is taken in the Stratonovich sense and $t_i$ is the starting time of $i^{\text{th}}$ cycle. Owing to its stochastic nature, $W_{cyc}$ of the engine fluctuates from cycle-to-cycle and we quantified the nature of these fluctuations through the probability distribution function  $\rho(W_{cyc})$. Figure \ref{Figure2}a and b show $\rho(W_{cyc})$ at different $\tau$ for the thermal and non-Gaussian Stirling cycles, respectively. Focusing on the large cycle duration ($\tau = 32$ s) first, we observed that $\rho(W_{cyc})$ is a Gaussian for the thermal and also for the non-Gaussian cycles (circles in Figure \ref{Figure2}a and b). The experimentally calculated average work done per cycle, $\langle W_{cyc} \rangle$, is negative indicating that the engine extracts heat from the bath to perform work on the surroundings. Further, $\tau = 32$ s corresponds to the quasistatic limit for both since the value of $\langle W_{cyc} \rangle$ is in excellent agreement with the theoretically calculated quasistatic Stirling work output, $W_{\infty} = k_B (T_{eff}^C - T_{eff}^H) \ln  \sqrt{\frac{k_{max}}{k_{min}}}$ (short solid horizontal lines in Figure \ref{Figure2}c).  

On lowering $\tau$, $\rho(W_{cyc})$ for the thermal Stirling engine remained a Gaussian (Figure \ref{Figure2}a) and $\langle W_{cyc}(\tau)\rangle \approx \langle W_{cyc}(\tau = 32 \text{ s})\rangle$ (hollow circles Figure \ref{Figure2}c). As expected of such a distribution, $\langle W_{cyc}\rangle$ was the same as the most-probable work $W^*$ - the value of $W_{cyc}$ where $\rho(W_{cyc})$ is a maximum (solid circles Figure \ref{Figure2}c). For the non-Gaussian engine on the other hand, on reducing $\tau$, $\rho(W_{cyc})$ became increasingly negatively skewed (Figure \ref{Figure2}b) and  $W^*(\tau)$ also became increasingly positive (solid squares Figure \ref{Figure2}c). $\langle W_{cyc}(\tau)\rangle$ however, was marginally smaller than $\langle W_{cyc}(\tau = 32 \text{ s})\rangle$. (hollow squares Figure \ref{Figure2}c).  We note that the work done by a thermal Stirling engine at a finite $\tau$ is given by the empirical relation \cite{schmiedl2007efficiency,blickle2012realization}
\begin{equation}
W(\tau) = W_{\infty} + W_{diss} \equiv W_{\infty} + \frac{\Sigma}{\tau}
\label{1}
\end{equation}
where, $W_{diss}$ is the dissipative work which accounts for the particle's inability to fully explore the available phase space when $k$ is rapidly lowered during the hot isotherm and $\Sigma$ is a constant also called the irreversibility parameter. Since $W_{diss}$ is a positive quantity as per definition, at small enough $\tau$, the overall work done itself can be positive indicating the  stalling of the engine. Clearly there is no buildup of irreversibility for the thermal engine as $\tau$ is lowered since $\langle W_{cyc}(\tau) \rangle \equiv W^*(\tau) \approx W_\infty$, while for the non-Gaussian one, there is, even if only in the most-probable sense ($\langle W_{cyc}(\tau) \rangle \approx W_\infty < W^*(\tau)$), and the engine stalls for $\tau\leq 10$ s. We also found excellent agreement between equation (\ref{1}) and our data allowing us to determine $\Sigma = 0.11\text{ }k_BT_{eff}^C$ (red solid line in Figure \ref{Figure2}c).

The observed behaviour of the non-Gaussian engine can be easily rationalized by analyzing the relaxation of the particle in the hot isotherm at the level of an individual cycle. For the particle to fully sample the statistical properties of the non-Gaussian hot reservoir, it should also experience the occasional large kicks that displaces it far from the center and not just the ones that predominantly keep it confined close to it. As $\tau$ is lowered, in \textit{most} cycles, the probability that the particle encounters a large kick in the isothermal expansion step also becomes increasingly small. Due to the incomplete exploration of the available phase volume in these cycles, less useful work is performed and $W^*(\tau)$ lifts off with decreasing $\tau$. In a few cycles, where these large kicks are present, anomalously large work is done by the engine and this results in $\rho(W_{cyc})$ being negatively skewed. When an adequate number of cycles, which has to be increased when $\tau$ is lowered, have been performed, all features of the noise are sampled and the engine operates like one in the quasistatic limit in an average sense with $\langle W_{cyc}(\tau)\rangle \to W_\infty$ (Figure \ref{Figure2}c). This inference can be strengthened by quantifying the equilibration of the particle over a fixed, but limited, number of cycles for all $\tau$. In Figure \ref{Figure2}d, we show ${k\langle x^2\rangle\over k_BT_{eff}^H}$ calculated over a small window in the middle of the hot isotherm and \textit{averaged} over $N = 50$ cycles for the thermal (squares) and the non-Gaussian engine (circles). Despite $N$ being small, ${k\langle x^2\rangle\over k_BT_{eff}^H}$ is close to 1 at all $\tau$ for the thermal engine implying that it is truly in the quasistatic limit, while for the non-Gaussian engine this is the case only at large $\tau$ with a clear violation of quasistaticity setting in for $\tau \leq 10$ s. Evidently, for a non-Gaussian engine $W^*(\tau)$, and not $\langle W_{cyc} (\tau) \rangle$, is a more precise metric for performance.

\subsection*{Tuning the performance of a Stirling engine through \textit{memoryless} non-Gaussian noise}

\begin{figure}[tbp]
\includegraphics[width=0.64\textwidth]{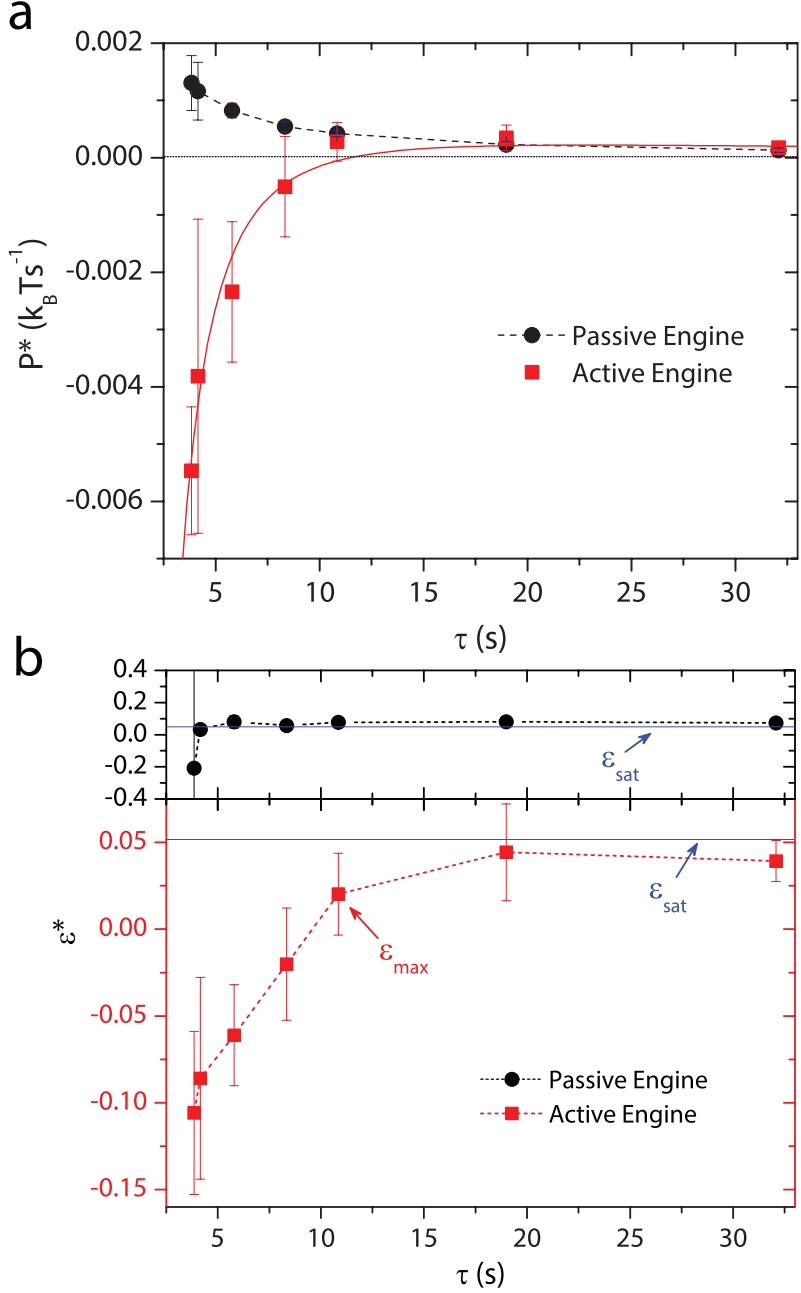}
\caption{\textbf{Quantifying the performance of a non-Gaussian Stirling engine.} In \textbf{a}, red squares (black circles) show the most-probable power $P^*$ of the non-Gaussian (Gaussian) engine at various $\tau$. $P^*$ increases slightly and then rapidly falls for the non-Gaussian engine for $\tau \leq 10.6$ s. The red solid line is calculated from the fit to Equation \ref{1} and is overlaid on the experimental data. \textbf{b} Red squares (black circles) represent the most-probable efficiency $\varepsilon^*$ of the non-Gaussian (Gaussian) engine at various $\tau$. The blue solid lines indicate the theoretically calculated saturation Stirling saturation, $\varepsilon_{Sat}$.  Efficiency $\varepsilon _{max} $ just before the rapid drop in power ($\tau = 10.6$) of the non-Gaussian engine agrees with the Curzon-Ahlborn efficiency $\varepsilon_{CA}$. Note that the black vertical line through the first data point (smallest $\tau$) is a portion of a large error bar. The error bars at other $\tau$ values are smaller than the symbol size.}
\label{Figure3}
\end{figure}

We now examined how differences in the nature of noise-statistics affected the power output of our engines. In the quasistatic limit $P(\tau) = -{\langle W_{cyc}(\tau)\rangle\over\tau} \to 0$ since $\tau\to\infty$, while at high cycle frequencies $W_{diss}$ is large and $P$ is once again small. At intermediate $\tau$, however, these effects compete resulting in a maximum in $P$ and this is a feature of both macroscopic and mesoscopic engines \cite{blickle2012realization,curzon1975efficiency}.  Figure \ref{Figure3}a shows the most-probable power, $P^*(\tau) = {-W^*(\tau)\over\tau}$, for the Gaussian Stirling engine (circles) and the non-Gaussian one (squares). Since for the thermal engine, over the range of $\tau$ studied $\Sigma = 0$, $P^*(\tau)$, which is same as same as $P(\tau)$, only increases monotonically on lowering $\tau$ and does not exhibit a maximum. Whereas for the non-Gaussian engine, on reducing $\tau$, $P^*(\tau)$ first appears to increase slightly, crosses zero for $\tau < 10$ s and then becomes more negative indicating stalling of the engine. We emphasize that for a Stirling cycle executed under conditions identical to that in Figure \ref{Figure1}e but where the non-Gaussian reservoir is replaced by a Gaussian one with the same $T_{eff}^H$, the maximum in $P$ is expected to be at a $\tau$ that is lower than even the Gaussian engine studied here (see Supplementary Note 3). This clearly shows that, even sans memory, altering the statistical properties of the noise bath alone allows for tuning the performance characteristics of mesoscopic heat engines. 

For a complete understanding of the operation of the non-Gaussian engine, we calculated its efficiency at various $\tau$ and benchmarked it with the thermal engine. Conventionally, the efficiency, $\varepsilon = {W_{cyc}\over Q}$, where $Q$ is the heat absorbed by the particle when it is in contact with the hot reservoir. $Q$ is the sum of the isochoric heat during the transition from state point \textcircled{2} to \textcircled{3}, $Q_{2\to3} = -{1\over 2}k_{max}(T_{eff}^H - T_{eff}^C)$ and the isothermal heat during transition from \textcircled{3} to \textcircled{4}, $Q_{3\to4} = \int_{(3)}^{(4)} \frac{\partial U}{\partial x}\dot{x} dt = W_H + Q_{boundary}$. Here, $W_H = \frac{1}{2}\int_{(3)}^{(4)} x^2\circ dk$ is the work done in the hot isotherm and $Q_{boundary} = -\frac{1}{2}[k(t)x^2(t)]_{(3)}^{(4)}$. For the non-Gaussian engine, we naturally chose $W^*$ instead of $W_{cyc}$ and defined the most-probable efficiency $\varepsilon^* = {W^*\over {\langle W_{H}\rangle+\langle Q_{boundary}\rangle+\langle Q_{isochoric}\rangle}}$ (See Supplementary Note 4). For the thermal engine, the experimentally determined $\varepsilon^*$ (black circles in Figure \ref{Figure3}b) hovers around the theoretically calculated saturation Stirling efficiency $\varepsilon_{Sat} =\varepsilon_c[1+{\varepsilon_c \over \ln(k_{max}/k_{min})}]^{-1}$ (solid blue line). Here, $\varepsilon_c = 1-{T_{eff}^C\over T_{eff}^H}$ is the Carnot efficiency. Whereas for the non-Gaussian engine, $\varepsilon^*(\tau)$ converges to $\varepsilon_{Sat}$ only at large $\tau$ (red squares in Figure \ref{Figure3}a). When $\tau$ is reduced, $\varepsilon^*(\tau)$ drops and becomes negative for $\tau<10$ s indicating stalling of the engine. Of particular importance in the operation of real heat engines is the efficiency at maximum power $\varepsilon_{Max}$, which for the non-Gaussian engine is at $\tau = 10.3$ s with $\varepsilon_{Max} = 0.025$. Most remarkably, this value in excellent agreement with theoretically predicted Curzon-Ahlborn efficiency, $\varepsilon_{CA} = {\varepsilon_{Sat} \over 2-\alpha\varepsilon_{Sat}} = 0.026$ \cite{curzon1975efficiency,schmiedl2007efficiency}.  In our experiments, $\alpha \sim 0$ is a constant calculated from the irreversibility parameters corresponding to the work done in the hot and cold isotherms (Supplementary Figure 4 and Supplementary Note 5). While it is known that $\varepsilon_{Max}\approx\varepsilon_{CA}$ for both macro and mesoscopic thermal engines, ours is the first observation of this being the case even for a non-Gaussian engine. 

\section*{\large Discussion}
Collectively, our experiments show that a micrometer-sized Stirling engine operating between a Gaussian and a non-Gaussian bath, without memory, indeed performs like a conventional engine in the quasistatic limit as anticipated by theory. On lowering the cycle times, the buildup of irreversibility in the engine, due entirely to the non-Gaussian nature of noise, results in work distributions that become increasingly negatively skewed unlike a thermal engine where it remains Gaussian. Strikingly, this noise-induced enhancement of irreversibility modulates the performance characteristics of the non-Gaussian engine in a manner similar to predictions by Curzon and Ahlborn for thermal engines where irreversibility sets in purely due to the rapid change of the control parameter. Our experiments thus reveal a new strategy for optimizing the performance of a mesoscale engine by tuning only the nature of noise statistics. Importantly, the ease with which the noise can be engineered and also applied locally, i.e. on the particle scale, in our approach presents advantages over other reservoir engineering methods where this can prove to be difficult, if not impossible \cite{martinez2013effective,martinez2017colloidal}. This should now make feasible the experimental realization of new stochastic machines like the non-Gaussian and the Buttiker–Landauer ratchet \cite{luczka1997symmetric,buttiker1987transport,landauer1988motion}.

\section*{\large Methods}
\subsection*{Experimental set-up for Reservoir Engineering}
In order to impart additional noise into the trapped colloid, a secondary optical trap was flashed along a line passing through the time-averaged centre of the particle at variable distances from the same. This was achieved by coupling a second laser (Excelsior 1064 nm, Spectra Physics USA) to the microscope which is reflected from a Spatial Light Modulator (Boulder Nonlinear Systems USA). The Spatial Light Modulator (SLM) contains a $512 \times 512$ array of shiny electrodes covered with a transparent liquid crystal layer so that an electric potential modulation across the electrodes imposes an additional phase pattern on the incident beam. We interfaced the SLM to a computer so that a series of desired phase patterns can be fed to the SLM at a fixed frequency of $34 Hz$. This enabled us to dynamically reconfigure the position of the first order diffraction spot by applying a series of linear diffraction grating patterns with varying periodicity which is controlled through a computer. We blocked the zeroth order spot so that only the first order spot is incident on the back of the microscope objective resulting in a flickering optical   trap in the vicinity of the tweezed colloidal particle. 

\subsection*{Image acquisition and processing}
Images of the trapped colloid was captured at $250$ Hz using a fast camera (Photron 500K-M3) attached to the microscope. Position of particle's centre in each frame was located at the subpixel level using the particle tracking codes by R. Parthasarathy \cite{parthasarathy2012rapid}. This allowed us to find the particle's position within an accuracy of $5$ nm. 

\subsection*{Non-Gaussian Reservoir Engineering}
For engineering the non-Gaussian reservoir, $\delta a$ were chosen from a $\delta - correlated$ distribution with zero mean and skewness but an extremely high kurtosis of 50. One such distribution with standard deviation of $\sigma = 0.28 \mu m$ is represented in Supplementary Figure 2(b). To create this distribution, we first generate two highly asymmetric distributions $\delta a_L$ and $\delta a_R$ (Supplementary Figure 2(a)) with a standard deviation of $0.28 \mu m$, a kurtosis of $60$ and a skewness of $-6.5$ for $\delta a_L$ and $+6.5$ for $\delta a_R$ through Pearson's protocol in MATLAB. Next we add/subtract a suitable number to $\delta a_L$ and $\delta a_R$ so that their peaks coincide at zero. Then we take union of $\delta a_L$ and $\delta a_R$ and randomly permute all the elements to finally obtain the set of $\delta a$.  In order to realize a desired effective temperature with such a noise, the standard deviation of $\delta a$ is optimized. It should be noted that heavy tails rise due to extremely rare events that can only be captured with a huge statistics. Since we are limited by a flashing frequency of $34 Hz$, it is not possible to completely sample the statistics with in one isotherm even for the largest $\tau$. To address this issue, the engine was cycled enough number of times (depending on $\tau$) so that the collection of all the hot isotherms exhausts all the rare events.

\subsection*{Instantaneous isochoric transitions}
The isochoric transitions \ding{173}$\rightarrow$\ding{174} and \ding{175}$\rightarrow$\ding{172} shown in Figure 1e of the main text is realised by changing the statistics and the variance of $\delta a$-distribution. The transition \ding{173}$\rightarrow$\ding{174} is realised by changing the $\delta a$ distribution from a Gaussian resulting in  $T_{eff}=1570 K$ to a non-Gaussian producing $T_{eff} = 1824.3 K$ while the transition \ding{175}$\rightarrow$\ding{172} is realised by the reverse.  Since the secondary laser is diffracted by a computer controlled SLM, the distribution from which $\delta a$s are chosen can be altered in $1/34$th of a second. Thus the particle is decoupled and coupled from one engineered reservoir to the other in less than $33 ms$ which is almost negligible even in compared to the lowest cycle time and hence instantaneous. 

\section*{\large Acknowledgements}
N.R. thanks Dr. Sudeesh Krishnamurthy for fruitful discussions. N.R. thanks Jawaharlal Nehru Centre for Advanced Scientific Research (JNCASR) for financial support. AKS thanks Department of Science and Technology (DST), Govt. of India for a Year of Science Fellowship. RG thanks JNCASR for financial support.

\section*{\large Author Contributions}

N.R., N.L., A.K.S. and R.G. designed experiments. N.R., N.L. and R.G. devised experimental procedures. N.R. performed experiments and carried out data analysis. N.R. and R.G. wrote the paper with inputs from all N.L. and A.K.S. 

\newpage
\renewcommand{\refname}{References}
\bibliographystyle{naturemag}		
\bibliography{references}

\singlespacing

\end{document}